\documentclass[11pt]{article}

\usepackage[margin=1in]{geometry}
\usepackage{amsmath,amssymb,amsthm}
\usepackage{graphicx}
\usepackage[hidelinks]{hyperref}

\newtheorem{proposition}{Proposition}

\theoremstyle{remark}
\newtheorem{remark}{Remark}

\begin{document}

\title{Equilibria of Replicator Dynamics in Quantum Games}
\author{Azhar Iqbal\\
{\small School of Electrical and Mechanical Engineering, Adelaide University,}\\
{\small Adelaide, SA 5005, Australia}
\and
A. H. Toor\\
{\small Department of Physics, Quaid-i-Azam University,}\\
{\small Islamabad 45320, Pakistan}}
\date{}
\maketitle

\begin{abstract}
We analyse two-population replicator dynamics for asymmetric $2\times2$
games implemented with the Marinatto--Weber (MW) identity/bit-flip
protocol. The effective payoff matrices are explicit convex combinations
of relabelled classical matrices. Consequently, the dynamics depends on
the initial state only through its four computational-basis probabilities:
quantum phases, coherences and entanglement cannot be distinguished from
classical shared randomness by the payoffs or trajectories in this
protocol. At each pure-strategy vertex, the local eigenvalues equal the
negatives of the corresponding unilateral payoff advantages; a vertex
is asymptotically stable precisely when it is a strict Nash equilibrium,
and thus an evolutionarily stable strategy pair in the asymmetric-game
sense. Explicit examples show how varying the initial distribution
changes which vertex is stable. When an isolated interior rest point
exists, it is a saddle or a nonlinear centre according to the sign of
the product of the classical payoff cross-differences; the sign cannot
change under the MW mapping, although the point can enter or leave the
strategy square. We give all four vertex-stability conditions explicitly in a single
table, obtain a closed-form logistic solution on the parity-balanced
hyperplane $S=0$, classify the degenerate cases, and give a first
integral for the nondegenerate flow. All results concern the MW
payoff protocol and its classically simulable family of bimatrix games;
none constitutes evidence of an inherently quantum evolutionary advantage.
\end{abstract}

\section{Introduction}

The theory of quantum games \cite{Piotrowski,Piotrowski2,Flitney} has attracted
significant attention in recent years. One area in which classical
game theory \cite{Neumann} has been particularly successful is the study of
evolutionary dynamics in populations. Evolutionary game theory
\cite{Weibull,Smith} treats the individuals of a population as players in a
game. The players need not be rational: their strategies are inherited
traits. In the original work of Maynard Smith and Price \cite{Smith,MaynardSmith},
anonymous players are randomly matched in pairs to play a matrix game, and
allowing higher-payoff strategies gradually to displace lower-payoff ones
introduces a dynamics into the population. Much of evolutionary game theory
is built around the concept of an evolutionarily stable strategy (ESS):
a strategy such that, if all individuals adopt it, no rare mutant strategy
can invade.

In earlier work \cite{iqbal,iqbal1,iqbal2,iqbal3,iqbal4} we studied how the
evolutionary stability of Nash equilibria (NE) is affected by the
quantization of classical games, using mainly the scheme proposed by
Marinatto and Weber \cite{marinatto}. We found that in certain games the
evolutionary stability of a NE can change as the game switches between its
classical and quantum forms. In those papers the ESS was treated as a
static concept. The present paper is concerned exclusively with the
two-strategy ($2\times2$) MW scheme; the extension of that scheme to
three-strategy games introduced in Refs.~\cite{iqbal3,iqbal4} has been
discussed separately by Fr\c{a}ckiewicz \cite{Frackiewicz}, who identified
a property of the generalised scheme that does not arise in the $2\times2$
case studied here.

Maynard Smith and Price \cite{Smith} introduced the ESS essentially as a
static concept, and nothing in the definition guarantees that evolution in
small mutational steps will actually converge to an ESS \cite{cressman}.
A dynamic approach \cite{cressman} relates the ESS to the
classical notions of stability for dynamical systems. A rest point
$(p^{\star},q^{\star})$ of a dynamical system is \emph{Lyapunov stable}
if for every neighbourhood $U$ of $(p^{\star},q^{\star})$ there is a
neighbourhood $V\subseteq U$ such that every trajectory starting in $V$
remains in $U$ for all later times. It is \emph{asymptotically stable}
(an \emph{attractor}) if, in addition, there is a neighbourhood $W$ such
that every trajectory starting in $W$ converges to $(p^{\star},q^{\star})$.

Taylor and Jonker \cite{TaylorJonker} introduced a dynamics into
evolutionary games under the hypothesis that the growth rate of the
subpopulation playing a given strategy is proportional to the advantage of
that strategy. This is now known as the replicator dynamics, one of many
dynamics studied in evolutionary game theory \cite{cressman,Bomze,sigmund1}.
Let $x_{i}$ be the frequency of individuals using strategy $i$, and let
$\mathbf{x}=(x_{1},\dots,x_{n})^{T}$ with $\sum_{i=1}^{n}x_{i}=1$. Let
$P_{i}(\mathbf{x})$ be the payoff to strategy $i$ when the population is in
state $\mathbf{x}$, and $\bar{P}=\sum_{j}x_{j}P_{j}(\mathbf{x})$ the mean
payoff. The replicator equation is \cite{sigmund1}
\begin{equation}
\dot{x}_{i}=x_{i}\bigl(P_{i}(\mathbf{x})-\bar{P}\bigr), \label{rep eq}
\end{equation}
where the dot denotes the time derivative. If the game is given by a payoff
matrix $A=(a_{ij})$, with $a_{ij}$ the payoff to strategy $i$ against
strategy $j$, then under random pairwise encounters
$P_{i}(\mathbf{x})=(A\mathbf{x})_{i}=a_{i1}x_{1}+\dots+a_{in}x_{n}$ and
Eq.~(\ref{rep eq}) becomes
\begin{equation}
\dot{x}_{i}=x_{i}\bigl((A\mathbf{x})_{i}-\mathbf{x}^{T}A\mathbf{x}\bigr).
\label{rep eq1}
\end{equation}
The population state is a point of the $(n-1)$-dimensional simplex
$\Delta$ \cite{Zeeman}, and Eq.~(\ref{rep eq1}) defines a flow on $\Delta$
whose orbits describe the evolution of the population. For such symmetric,
single-population games every ESS is an asymptotically stable rest point of
Eq.~(\ref{rep eq1}), but the converse does not hold: an attractor need not
be an ESS \cite{TaylorJonker,HSS,Zeeman,Weibull}.

For asymmetric games played between two distinct populations (for example
males and females) the situation is different. Selten \cite{Selten} showed
that in such games an ESS must be a strict NE, and for the two-population
replicator dynamics a strategy pair is asymptotically stable if and only if
it is a strict NE \cite{RitzbergerWeibull,sigmund1,HS2003,CressmanTao}. The static and
dynamic notions therefore coincide.

In this paper we ask how the attractors of the replicator dynamics of an
asymmetric $2\times2$ game, and their evolutionary stability, respond to
quantization in the Marinatto--Weber scheme. Our main results are as
follows. (i)~Because the quantum game is itself a bi-matrix game, the
coincidence of ESSs and attractors survives the MW transformation: at every vertex
of the state space, the eigenvalues of the linearized replicator dynamics
equal the payoff losses from unilateral deviation
(Proposition~\ref{prop:vertex}); the complete set of stability conditions
for all four vertices is collected in Table~(\ref{eq:vertex-table}).
(ii)~Quantization can nevertheless change
\emph{which} strategy pairs are ESS attractors: we give conditions on the
initial quantum state under which a saddle point that is not an ESS of the
classical game becomes an ESS attractor of the quantum game, and
conversely (Sec.~\ref{sec:cases}). (iii)~The type of an isolated interior rest
point (saddle or centre) is fixed by the classical payoff cross-differences;
the MW transformation can move it into or out of the state space
(Proposition~\ref{prop:interior}); on the parity-balanced hyperplane $S=0$
the dynamics decouples and admits an exact logistic solution
(Sec.~\ref{sec:balanced}). (iv)~The protocol is classically simulable using
shared randomness, so these stability changes do not establish an
entanglement-dependent advantage (Proposition~\ref{prop:simulation}).

\section{Replicator dynamics of a quantized bi-matrix game}

Schuster and Sigmund \cite{schuster1} applied the replicator dynamics to
asymmetric animal conflicts, deriving explicit rest-point conditions for
the Battle of the Sexes. Their framework provides the classical baseline
extended here to the MW payoff protocol. Consider the general bi-matrix game
\begin{equation}
\begin{array}{c}
\text{Male's strategy}
\end{array}
\begin{array}{c}
X_{1} \\
X_{2}
\end{array}
\overset{
\begin{array}{cc}
\text{Female's} & \text{strategy} \\
Y_{1} & Y_{2}
\end{array}
}{\left[
\begin{array}{cc}
(a_{11},b_{11}) & (a_{12},b_{12}) \\
(a_{21},b_{21}) & (a_{22},b_{22})
\end{array}
\right] }  \label{genMatrix}
\end{equation}
in which a male plays the pure strategies $X_{1}$, $X_{2}$, a female plays
the pure strategies $Y_{1}$, $Y_{2}$, and $(a_{ij},b_{ij})$ are the payoffs
to the male and the female, respectively, when the male plays $X_{i}$ and
the female plays $Y_{j}$. Let $x_{1},x_{2}$ be the frequencies of $X_{1}$
and $X_{2}$ in the male population and $y_{1},y_{2}$ the frequencies of
$Y_{1}$ and $Y_{2}$ in the female population, so that
\begin{equation}
x_{1}+x_{2}=y_{1}+y_{2}=1,\qquad x_{i}\geq 0,\ y_{i}\geq 0\quad(i=1,2).
\label{cond}
\end{equation}
When the male plays $X_{1}$ with probability $x$ and the female plays
$Y_{1}$ with probability $y$, the payoffs are \cite{Broom}
\begin{equation}
P_{M}(x,y)=\mathbf{x}^{T}\mathbf{M}\mathbf{y},\qquad
P_{F}(x,y)=\mathbf{x}^{T}\mathbf{F}\mathbf{y},  \label{payoffs}
\end{equation}
where
\begin{equation*}
\mathbf{M}=\begin{bmatrix} a_{11} & a_{12} \\ a_{21} & a_{22}\end{bmatrix},
\quad
\mathbf{F}=\begin{bmatrix} b_{11} & b_{12} \\ b_{21} & b_{22}\end{bmatrix},
\quad
\mathbf{x}=\begin{bmatrix} x \\ 1-x\end{bmatrix},
\quad
\mathbf{y}=\begin{bmatrix} y \\ 1-y\end{bmatrix}.
\end{equation*}
Writing $x_{1}=x$ and $y_{1}=y$, the two-population replicator dynamics
$\dot{x}=x(1-x)\,\partial P_{M}/\partial x$,
$\dot{y}=y(1-y)\,\partial P_{F}/\partial y$ reads
\begin{align}
\dot{x} &= x(1-x)\bigl\{ (a_{12}-a_{22})+y\,(a_{11}-a_{12}-a_{21}+a_{22})\bigr\},
\notag \\
\dot{y} &= y(1-y)\bigl\{ (b_{21}-b_{22})+x\,(b_{11}-b_{12}-b_{21}+b_{22})\bigr\}.
\label{RepEqs}
\end{align}
Note that the constant term in the female equation is $b_{21}-b_{22}$: when
all males play $X_{2}$ ($x=0$), a female gains $b_{21}$ from $Y_{1}$ and
$b_{22}$ from $Y_{2}$. Equations of this form are closely related to the
Lotka--Volterra equations of population ecology \cite{Hirsch,sigmund1}.

Adding a constant to a column of $\mathbf{M}$ (or to a row of $\mathbf{F}$)
leaves Eqs.~(\ref{RepEqs}), and all Nash and strict Nash conditions,
unchanged. Following Schuster and Sigmund \cite{schuster1} we can therefore
take, without loss of generality,
\begin{equation}
a_{11}=a_{22}=b_{11}=b_{22}=0,\qquad a_{12}=a,\quad a_{21}=b,\quad
b_{12}=c,\quad b_{21}=d.  \label{simply}
\end{equation}
The same normalization is also without loss of generality for the
quantum game below, because the quantization map (\ref{omega})--(\ref{chi})
sends a column shift of $\mathbf{M}$ to a column shift of $\boldsymbol{\omega}$
(and a row shift of $\mathbf{F}$ to a row shift of $\boldsymbol{\chi}$).

Suppose now that a quantum form of the game (\ref{genMatrix}) is played
using the scheme of Marinatto and Weber \cite{marinatto}. The players share
the initial state
\begin{equation}
\left|\psi_{\mathrm{ini}}\right\rangle=\sum_{i,j=1,2}c_{ij}\left|ij\right\rangle,
\qquad \sum_{i,j=1,2}\left|c_{ij}\right|^{2}=1,  \label{norm}
\end{equation}
where the label $1$ corresponds to the pure strategies $X_{1}$, $Y_{1}$ and
the label $2$ to $X_{2}$, $Y_{2}$; the amplitudes $c_{ij}$ are in general
complex. The male applies the identity $\hat{I}$ to his qubit with
probability $x$ and the spin-flip operator $\hat{\sigma}_{x}$ (which
interchanges $\left|1\right\rangle$ and $\left|2\right\rangle$) with
probability $1-x$; the female does the same with probabilities $y$ and
$1-y$. The payoffs are then determined by a measurement on the final
state \cite{marinatto}. They take the same form as Eq.~(\ref{payoffs})
\cite{iqbal4},
\begin{equation}
P_{M}(x,y)=\mathbf{x}^{T}\boldsymbol{\omega}\mathbf{y},\qquad
P_{F}(x,y)=\mathbf{x}^{T}\boldsymbol{\chi}\mathbf{y},
\end{equation}
where $\boldsymbol{\omega}=(\omega_{ij})$ and $\boldsymbol{\chi}=(\chi_{ij})$
are the quantum forms of $\mathbf{M}$ and $\mathbf{F}$:
\begin{align}
\omega_{11} &= a_{11}|c_{11}|^{2}+a_{12}|c_{12}|^{2}+a_{21}|c_{21}|^{2}+a_{22}|c_{22}|^{2}, \notag\\
\omega_{12} &= a_{11}|c_{12}|^{2}+a_{12}|c_{11}|^{2}+a_{21}|c_{22}|^{2}+a_{22}|c_{21}|^{2}, \notag\\
\omega_{21} &= a_{11}|c_{21}|^{2}+a_{12}|c_{22}|^{2}+a_{21}|c_{11}|^{2}+a_{22}|c_{12}|^{2}, \notag\\
\omega_{22} &= a_{11}|c_{22}|^{2}+a_{12}|c_{21}|^{2}+a_{21}|c_{12}|^{2}+a_{22}|c_{11}|^{2},
\label{omega}
\end{align}
and $\chi_{ij}$ is obtained from $\omega_{ij}$ by replacing each $a_{kl}$
with $b_{kl}$:
\begin{align}
\chi_{11} &= b_{11}|c_{11}|^{2}+b_{12}|c_{12}|^{2}+b_{21}|c_{21}|^{2}+b_{22}|c_{22}|^{2}, \notag\\
\chi_{12} &= b_{11}|c_{12}|^{2}+b_{12}|c_{11}|^{2}+b_{21}|c_{22}|^{2}+b_{22}|c_{21}|^{2}, \notag\\
\chi_{21} &= b_{11}|c_{21}|^{2}+b_{12}|c_{22}|^{2}+b_{21}|c_{11}|^{2}+b_{22}|c_{12}|^{2}, \notag\\
\chi_{22} &= b_{11}|c_{22}|^{2}+b_{12}|c_{21}|^{2}+b_{21}|c_{12}|^{2}+b_{22}|c_{11}|^{2}.
\label{chi}
\end{align}
For the product state $\left|\psi_{\mathrm{ini}}\right\rangle=\left|11\right\rangle$
we have $\boldsymbol{\omega}=\mathbf{M}$ and $\boldsymbol{\chi}=\mathbf{F}$,
so the classical game is embedded in the quantum one. More generally,
for any basis state $|ij\rangle$ ($i,j\in\{1,2\}$) the matrices
$\boldsymbol{\omega}$ and $\boldsymbol{\chi}$ are permutations of
$\mathbf{M}$ and $\mathbf{F}$ respectively, so the classical game is
recoverable (up to a relabelling of strategies) from any product basis
initial state. This is a structural feature of the $2\times2$ MW scheme
that does not extend automatically to larger strategy spaces
\cite{Frackiewicz}. Throughout, the
MW game is the classical bi-matrix game with payoff matrices
$(\boldsymbol{\omega},\boldsymbol{\chi})$, and Eqs.~(\ref{RepEqs}) apply with
$a_{ij}\to\omega_{ij}$, $b_{ij}\to\chi_{ij}$.

\begin{proposition}[Classical simulation of the MW payoff dynamics]
\label{prop:simulation}
Let $p_{ij}=|c_{ij}|^2$ and let $R$ and $C$ denote the two-by-two
row- and column-swap permutation matrices. Then
\begin{equation}
\boldsymbol{\omega}=p_{11}\mathbf M+p_{12}\mathbf M C
+p_{21}R\mathbf M+p_{22}R\mathbf M C,\qquad
\boldsymbol{\chi}=p_{11}\mathbf F+p_{12}\mathbf F C
+p_{21}R\mathbf F+p_{22}R\mathbf F C.
\label{eq:convex}
\end{equation}
For every pure initial state, the same expected payoffs and replicator
trajectories arise from the separable, classically correlated density
matrix $\rho_{\rm cl}=\sum_{ij}p_{ij}|ij\rangle\langle ij|$.
More generally, every probability vector $p_{ij}\geq0$ with
$\sum p_{ij}=1$ is obtainable from either such a diagonal classical
mixture or a pure state $\sum\sqrt{p_{ij}}|ij\rangle$.
Thus the protocol cannot distinguish entanglement or relative phases by
payoff or dynamical observations alone.
\end{proposition}
\begin{proof}
Condition on the basis pair $ij$ drawn with probability $p_{ij}=|c_{ij}|^2$.
Applying $\hat I$ or $\hat\sigma_x$ to a qubit in state $|i\rangle$
permutes its label, mapping $|ij\rangle$ to one of the four basis states.
The payoff for any pure choice of the two local operators is therefore
the corresponding entry of $\mathbf M$ (or $\mathbf F$). Since the
measurement observable is diagonal in the computational basis, only the
diagonal elements $p_{ij}$ of the initial density matrix contribute;
relative phases between basis components of $|\psi_{\mathrm{ini}}\rangle$
never appear. Averaging over the four basis-pair outcomes
yields Eq.~(\ref{eq:convex}). Mixing over the players' independent
classical choices $x,y$ is separately linear in those probabilities,
so the replicator vector field inherits the same independence of phases.
\end{proof}
\begin{remark}
Equation~(\ref{eq:convex}) describes an \emph{effective-game
transformation}, not a quantum effect irreducible to classical
randomness. The same correlated draw is shared between the two
players, and the mixing probabilities $x,y$ remain classical population
frequencies. No claim is made for protocols with a different operator
set or non-diagonal measurements. Related classical-correlation
interpretations of quantum games are discussed in
Refs.~\cite{Brandenburger,Frackiewicz2020}.
\end{remark}

Under the normalization
(\ref{simply}),
\begin{align}
\omega_{11} &= a|c_{12}|^{2}+b|c_{21}|^{2}, & \omega_{12} &= a|c_{11}|^{2}+b|c_{22}|^{2}, \notag\\
\omega_{21} &= a|c_{22}|^{2}+b|c_{11}|^{2}, & \omega_{22} &= a|c_{21}|^{2}+b|c_{12}|^{2}, \notag\\
\chi_{11} &= c|c_{12}|^{2}+d|c_{21}|^{2}, & \chi_{12} &= c|c_{11}|^{2}+d|c_{22}|^{2}, \notag\\
\chi_{21} &= c|c_{22}|^{2}+d|c_{11}|^{2}, & \chi_{22} &= c|c_{21}|^{2}+d|c_{12}|^{2}.
\end{align}
It is convenient to introduce
\begin{equation}
K_{1}=|c_{11}|^{2}-|c_{21}|^{2},\quad K_{2}=|c_{22}|^{2}-|c_{12}|^{2},\quad
L_{1}=|c_{11}|^{2}-|c_{12}|^{2},\quad L_{2}=|c_{22}|^{2}-|c_{21}|^{2},
\label{KL}
\end{equation}
and $S=K_{1}+K_{2}=L_{1}+L_{2}=|c_{11}|^{2}+|c_{22}|^{2}-|c_{12}|^{2}-|c_{21}|^{2}$.
Note that $K_{i}$ involves differences of same-column basis probabilities (associated
with the male's qubit) whereas $L_{i}$ involves differences of same-row basis
probabilities (associated with the female's qubit). The common sum
$S=|c_{11}|^{2}+|c_{22}|^{2}-|c_{12}|^{2}-|c_{21}|^{2}$ is the signed excess of
diagonal over off-diagonal basis probabilities; it depends only on the measurement
probabilities $p_{ij}=|c_{ij}|^{2}$, not on quantum phases or entanglement
(for instance, the maximally entangled Bell state
$(|11\rangle+|22\rangle)/\sqrt{2}$ gives $S=1$, the same value as the
product state $|11\rangle$).
The replicator equations (\ref{RepEqs}) of the quantum game then take the form
\begin{align}
\dot{x} &= x(1-x)\bigl[aK_{1}+bK_{2}-(a+b)S\,y\bigr], \notag\\
\dot{y} &= y(1-y)\bigl[cL_{2}+dL_{1}-(c+d)S\,x\bigr].
\label{QRpEs}
\end{align}
For $\left|\psi_{\mathrm{ini}}\right\rangle=\left|11\right\rangle$ we have
$K_{1}=L_{1}=S=1$ and $K_{2}=L_{2}=0$, and Eqs.~(\ref{QRpEs}) reduce to the
classical equations $\dot{x}=x(1-x)[a-(a+b)y]$, $\dot{y}=y(1-y)[d-(c+d)x]$.
The male and female equations involve different combinations of the
amplitudes ($K_{i}$ and $L_{i}$, respectively); this reflects the fact that
the male's strategy acts on the first qubit and the female's on the second.

The dynamics (\ref{QRpEs}) always has the four vertex rest points
$(0,0)$, $(0,1)$, $(1,0)$, $(1,1)$ and, provided $(a+b)(c+d)S\neq0$, the candidate rest point
\begin{equation}
x^{\ast}=\frac{cL_{2}+dL_{1}}{(c+d)S},\qquad
y^{\ast}=\frac{aK_{1}+bK_{2}}{(a+b)S},  \label{IntEq}
\end{equation}
which is an interior rest point when $0<x^{\ast},y^{\ast}<1$. For
$\left|\psi_{\mathrm{ini}}\right\rangle=\left|11\right\rangle$ it reduces to
the classical point $x^{\ast}=d/(c+d)$, $y^{\ast}=a/(a+b)$ \cite{schuster1}.

We use linearization \cite{Hirsch} to determine the local phase portrait.
Writing Eqs.~(\ref{QRpEs}) as $\dot{x}=\mathbf{X}(x,y)$,
$\dot{y}=\mathbf{Y}(x,y)$, the Jacobian matrix is
\begin{equation}
J=\begin{bmatrix}\mathbf{X}_{x} & \mathbf{X}_{y}\\ \mathbf{Y}_{x} & \mathbf{Y}_{y}\end{bmatrix},
\label{linztn}
\end{equation}
where, for example, $\mathbf{X}_{x}=\partial\mathbf{X}/\partial x$, and
\begin{align}
\mathbf{X}_{x} &= (1-2x)\bigl\{aK_{1}+bK_{2}-(a+b)S\,y\bigr\}, &
\mathbf{X}_{y} &= -x(1-x)(a+b)S, \notag\\
\mathbf{Y}_{x} &= -y(1-y)(c+d)S, &
\mathbf{Y}_{y} &= (1-2y)\bigl\{cL_{2}+dL_{1}-(c+d)S\,x\bigr\}.
\label{TermML}
\end{align}
The eigenvalues $\lambda_{1},\lambda_{2}$ of $J$ at a rest point are the
roots of the characteristic equation
\begin{equation}
\det(J-\lambda I)=0.  \label{charEq}
\end{equation}
For a hyperbolic rest point the standard planar classifications are: (1)~$\lambda_{1},\lambda_{2}$
real and negative: a stable node (attractor); real and positive: an
unstable node (repellor). (2)~$\lambda_{1},\lambda_{2}$ real and of opposite
sign: a saddle. (3)~$\lambda_{1,2}=\alpha\pm i\beta$ with $\beta\neq0$: a
stable spiral (attractor) if $\alpha<0$ and an unstable spiral (repellor)
if $\alpha>0$. If $\alpha=0$, the linearization alone cannot distinguish a centre
from a weak focus; for the system studied here a first integral resolves
this ambiguity (Remark~\ref{rem:centre}).

\section{Evolutionary stability and attractors at the vertices}

At a vertex the off-diagonal entries of $J$ vanish, so the eigenvalues are
the diagonal entries $\mathbf{X}_{x}$ and $\mathbf{Y}_{y}$. By symmetry of the
labelling it suffices to consider the vertex $x=1$, $y=0$, written $(1,0)$,
at which the male plays $X_{1}$ and the female plays $Y_{2}$. There
\begin{equation}
\lambda_{1}=-(aK_{1}+bK_{2}),\qquad \lambda_{2}=-(cL_{1}+dL_{2}).
\label{roots}
\end{equation}
For the classical game, $\lambda_{1}=-a$ and $\lambda_{2}=-c$.

Since the game (\ref{genMatrix}) is asymmetric, a strategy pair is an ESS
if and only if it is a strict NE \cite{Selten,Weibull}. The strict NE
conditions for $(1,0)$ in the quantum game are
\begin{align}
P_{M}(1,0)-P_{M}(x,0) &= (1-x)\bigl\{a\bigl(|c_{11}|^{2}-|c_{21}|^{2}\bigr)
+b\bigl(|c_{22}|^{2}-|c_{12}|^{2}\bigr)\bigr\}
=(1-x)(aK_{1}+bK_{2})>0, \notag\\
P_{F}(1,0)-P_{F}(1,y) &= y\bigl\{c\bigl(|c_{11}|^{2}-|c_{12}|^{2}\bigr)
+d\bigl(|c_{22}|^{2}-|c_{21}|^{2}\bigr)\bigr\}
=y(cL_{1}+dL_{2})>0,
\label{StNE}
\end{align}
for all $x,y\in[0,1]$ with $x\neq1$ and $y\neq0$,
respectively. In the classical game,
$(1,0)$ is therefore an ESS if and only if $a>0$ and $c>0$.

Comparing Eqs.~(\ref{roots}) and (\ref{StNE}) gives the following result.

\begin{proposition}\label{prop:vertex}
For every initial state $\left|\psi_{\mathrm{ini}}\right\rangle$, the
eigenvalues of the linearized replicator dynamics at the vertex $(1,0)$
are the negatives of the payoff advantages that $X_{1}$ and $Y_{2}$ have
over the alternative pure strategies,
\begin{equation*}
\lambda_{1}=-\bigl(\omega_{12}-\omega_{22}\bigr),\qquad
\lambda_{2}=-\bigl(\chi_{12}-\chi_{11}\bigr).
\end{equation*}
Consequently $(1,0)$ is a hyperbolic attractor if and only if it is a
strict NE, that is, if and only if it is an ESS of the quantum game.
The same computation applies at the other three vertices. Moreover,
when an eigenvalue is zero (a payoff advantage vanishes), the adjacent
boundary edge consists of stationary points and asymptotic stability
fails; this extends the biconditional to all vertices, including
non-hyperbolic ones (see Section~\ref{sec:degenerate}).
\end{proposition}

\begin{proof}
At $(1,0)$ the Jacobian is diagonal. From Eq.~(\ref{TermML}),
substituting $x=1$ gives $(1-2x)=-1$, so
$\mathbf{X}_{x}=-(aK_{1}+bK_{2})=-(\omega_{12}-\omega_{22})$,
since the male bracket evaluates to $\omega_{12}-\omega_{22}$ at $y=0$.
Similarly, substituting $x=1$ and $y=0$,
$\mathbf{Y}_{y}=(1-0)\{cL_{2}+dL_{1}-(c+d)S\}=-(cL_{1}+dL_{2})
=-(\chi_{12}-\chi_{11})$, because the female bracket at $x=1$ is
$\chi_{11}-\chi_{12}$. By Eq.~(\ref{StNE}) these are exactly minus the
coefficients in the two strict NE conditions, so both eigenvalues are
negative if and only if both strict inequalities hold.
\end{proof}

Proposition~\ref{prop:vertex} is the quantum counterpart of the classical
result that, for two-population replicator dynamics, asymptotically stable
rest points are exactly the strict NE
\cite{RitzbergerWeibull,sigmund1,HS2003,CressmanTao}.
It holds because the Marinatto--Weber quantum game is itself a bi-matrix
game. In particular, in this scheme a vertex can never be an attractor
without also being an ESS, or the reverse. What quantization \emph{can} change
is \emph{which} vertices are ESS attractors, and this is what we
illustrate next.

\subsection{Stability regions in the probability simplex}
\label{sec:chambers}
The initial distribution belongs to the three-dimensional simplex
$\Delta_3=\{p\in\mathbb R_{\geq0}^{4}:\sum_{ij}p_{ij}=1\}$.
The dependence of every vertex stability test on $p$ is affine. This
provides a global geometric description of the changes illustrated below.

\begin{proposition}[Polyhedral stability regions]
\label{prop:regions}
For a fixed classical bimatrix game and a fixed pure-strategy vertex
$v=(x_v,y_v)$, there exist two explicitly computable vectors
$g_M(v),g_F(v)\in\mathbb R^4$, depending only on the original payoffs,
for which $v$ is an asymptotically stable equilibrium of the
MW two-population replicator dynamics exactly when
\begin{equation}
 p\cdot g_M(v)>0,\qquad p\cdot g_F(v)>0.
 \label{eq:chambers}
\end{equation}
Thus the set of initial distributions giving stability at $v$ is the
intersection of the simplex with two open half-spaces, and is convex
(possibly empty). Its boundaries are contained in the two hyperplanes
$p\cdot g_M(v)=0$ and $p\cdot g_F(v)=0$. The full pattern of strict
vertex equilibria is constant on each connected open cell of the
finite hyperplane arrangement given by these payoff differences.
\end{proposition}
\begin{proof}
For each basis input $|ij\rangle$ and each vertex $v$, compute the
payoff advantage of the strategy played by the male over his alternative
with the female fixed at $y_v$; assign this to the $ij$ component of
$g_M(v)$. Define $g_F(v)$ analogously for the female with $x_v$
fixed. Formula~(\ref{eq:convex}) implies that the two unilateral
advantages are respectively $p\cdot g_M(v)$ and $p\cdot g_F(v)$.
Proposition~\ref{prop:vertex} then gives the strict inequalities
(\ref{eq:chambers}). Affinity proves the asserted half-space and
hyperplane properties.\qedhere
\end{proof}

For example, at $v=(1,0)$, with $p=(p_{11},p_{12},p_{21},p_{22})$,
\begin{align}
 g_M(1,0)&=(a,-b,-a,b),\notag\\
 g_F(1,0)&=(c,-c,-d,d),
 \label{eq:explicitg}
\end{align}
so the conditions are $aK_1+bK_2>0$ and $cL_1+dL_2>0$.
For completeness, write the effective payoff differences as
$A=aK_1+bK_2$, $B=(a+b)S$, $C=cL_2+dL_1$ and $D=(c+d)S$.
The following table lists the two \emph{resident-versus-mutant} payoff
advantages at every vertex. Both entries must be strictly positive for
that vertex to be an ESS attractor:
\begin{equation}
\begin{array}{c|cc}
\text{vertex }(x,y)&\text{male advantage}&\text{female advantage}\\hline
(0,0)&-A&-C\\
(1,0)& A&D-C\\
(0,1)& B-A&C\\
(1,1)& A-B&C-D
\end{array}.
\label{eq:vertex-table}
\end{equation}
This supplies every stability condition explicitly without enumerating
all eight payoff inequalities separately. Each entry is the
\emph{resident-minus-mutant} unilateral payoff advantage at the named vertex:
the sign of $A$ reverses at $(0,0)$ and $(0,1)$ because the male resident
plays $X_2$ there, and the sign of $C$ reverses at $(0,0)$ and $(1,1)$
because the female resident plays $Y_2$ there.
No additional quantum-state invariants beyond $A,B,C,D$ are required.

In particular, interpolating between two initial distributions
$p(t)=(1-t)p^{(0)}+tp^{(1)}$ makes each unilateral advantage linear in
$t$. It can switch sign only at a hyperplane crossing. At a crossing
where a single advantage vanishes, an adjacent boundary edge consists
of stationary points, as detailed in Sec.~\ref{sec:degenerate}.

The same description provides a robustness bound. If a vertex has
positive payoff advantages $m_M,m_F$ at $p$, then for any other
probability vector $q$,
\begin{equation}
|(q-p)\cdot g_k(v)|\leq\frac12\bigl(\max_{ij}g_{k,ij}(v)
-\min_{ij}g_{k,ij}(v)\bigr)\lVert q-p\rVert_1,
\quad k\in\{M,F\}.
\label{eq:robust}
\end{equation}
Consequently both strict advantages, and hence stability, persist
whenever the right-hand side is smaller than $m_k$ for each player.
This quantifies how far the shared initial distribution can be
perturbed without changing the identity of the stable vertex.

\subsection{Quantization can create and destroy ESS attractors}
\label{sec:cases}

\paragraph{Case (a).}
Let $a,d>0$ and $b,c<0$. In the classical game $(1,0)$ has
$\lambda_{1}=-a<0$ and $\lambda_{2}=-c>0$: it is a saddle and, since
$c<0$, not an ESS. Now choose the initial state so that
\begin{equation}
|c_{21}|^{2}<|c_{22}|^{2}<|c_{11}|^{2}<|c_{12}|^{2}.
\label{orderA}
\end{equation}
Then $K_{1}>0$, $K_{2}<0$, $L_{1}<0$, $L_{2}>0$, so that $aK_{1}+bK_{2}>0$ and
$cL_{1}+dL_{2}>0$. By Proposition~\ref{prop:vertex}, $(1,0)$ is both an ESS
and an attractor of the quantum game. Quantization has thus converted a
saddle that is not an ESS into an ESS attractor.

\paragraph{Case (b).}
Let $a,c,d>0$ and $b<0$. In the classical game $(1,0)$ has
$\lambda_{1}=-a<0$, $\lambda_{2}=-c<0$: it is an ESS attractor. Now choose
\begin{equation}
|c_{22}|^{2}<|c_{21}|^{2}<|c_{11}|^{2}<|c_{12}|^{2}.
\label{orderB}
\end{equation}
Then $K_{1}>0$, $K_{2}<0$, $L_{1}<0$, $L_{2}<0$, so that $aK_{1}+bK_{2}>0$
while $cL_{1}+dL_{2}<0$. The vertex $(1,0)$ is now a saddle and is no longer
an ESS of the quantum game.

\begin{figure}[!htbp]
\centering
\includegraphics[width=0.72\textwidth]{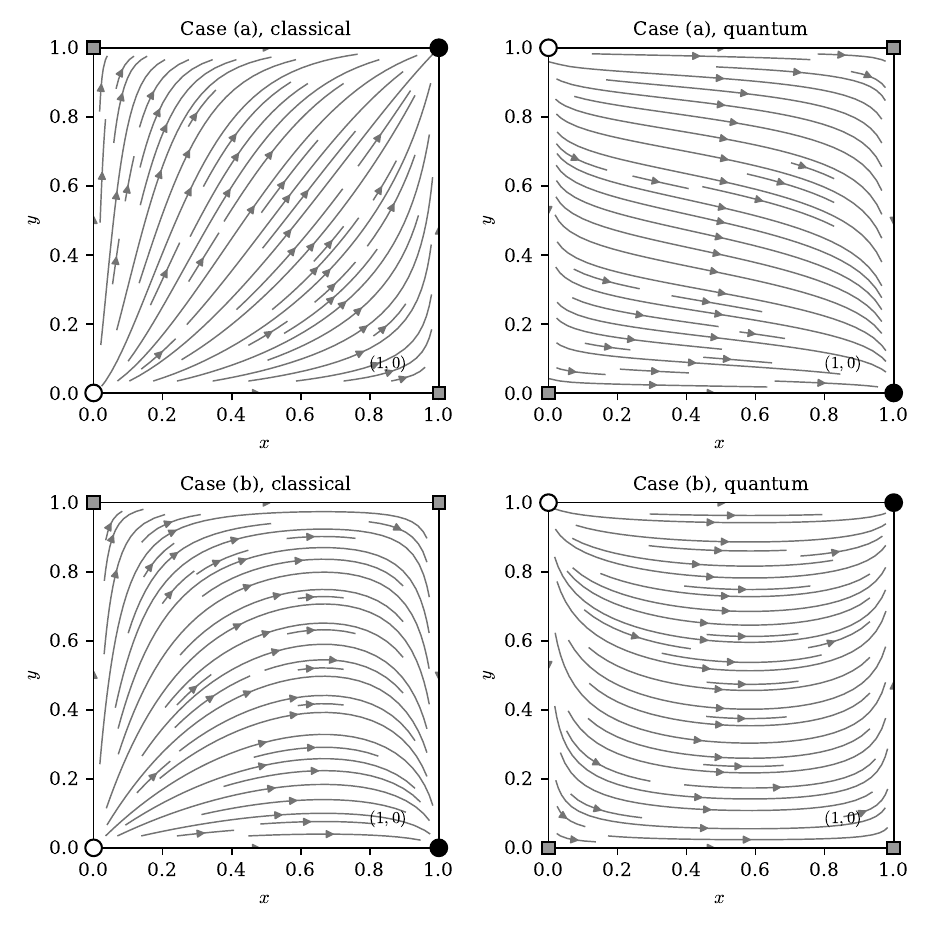}
\caption{Phase portraits of the replicator dynamics (\ref{QRpEs}) for the
classical game (left) and the Marinatto--Weber quantum game (right).
Top row, case~(a): $(a,b,c,d)=(2,-1,-1,3)$ and
$(|c_{11}|^{2},|c_{12}|^{2},|c_{21}|^{2},|c_{22}|^{2})=(0.36,0.37,0.10,0.17)$.
Bottom row, case~(b): $(a,b,c,d)=(2,-1,1,2)$ and
$(|c_{11}|^{2},|c_{12}|^{2},|c_{21}|^{2},|c_{22}|^{2})=(0.30,0.45,0.15,0.10)$.
Filled circles: attractors (which are also ESSs); open circles: repellors;
grey squares: saddles. In both cases quantization moves the ESS attractor
between the vertices $(1,1)$ and $(1,0)$.}
\label{fig:phase}
\end{figure}

Figure~\ref{fig:phase} shows the phase portraits for the concrete choices
$(a,b,c,d)=(2,-1,-1,3)$ with
$(|c_{11}|^{2},|c_{12}|^{2},|c_{21}|^{2},|c_{22}|^{2})=(0.36,0.37,0.10,0.17)$
for case~(a), and $(a,b,c,d)=(2,-1,1,2)$ with
$(|c_{11}|^{2},|c_{12}|^{2},|c_{21}|^{2},|c_{22}|^{2})=(0.30,0.45,0.15,0.10)$
for case~(b). In case~(a) the eigenvalues at $(1,0)$ change from
$(-2,\,1)$ to $(-0.72,\,-0.22)$; in case~(b) they change from $(-2,\,-1)$
to $(-0.65,\,0.25)$. In both examples quantization does not simply add or
remove an ESS: it moves the ESS attractor between the vertices $(1,1)$ and $(1,0)$
for these parameter choices.

A convenient way to locate the transitions is to interpolate between
the classical distribution $p^{(0)}=(1,0,0,0)$ and the displayed
distribution $p^{(1)}$ for each example, taking
$p(t)=(1-t)p^{(0)}+tp^{(1)}$ with $0\leq t\leq1$.
In case~(a), the female's advantage at $(1,0)$ is
$-1+1.22t$, whereas at $(1,1)$ it is $1-1.22t$.
The two vertices exchange stability at $t=50/61\simeq0.8197$.
In case~(b), the corresponding advantages are $1-1.25t$
and $-1+1.25t$, so the exchange occurs at $t=4/5$.
The male's advantages at the two relevant vertices remain positive
throughout each interpolation. At the threshold the edge $x=1$ is
stationary rather than providing a hyperbolic equilibrium. These
calculations specify the bifurcation in the probability parameter
without attributing it to quantum coherence.

\section{The interior rest point}

At the rest point (\ref{IntEq}) we have $\mathbf{X}_{x}=\mathbf{Y}_{y}=0$, and
\begin{align}
\mathbf{X}_{y} &= -x^{\ast}(1-x^{\ast})(a+b)S
=-\frac{(cL_{2}+dL_{1})(cL_{1}+dL_{2})(a+b)}{(c+d)^{2}S}, \notag\\
\mathbf{Y}_{x} &= -y^{\ast}(1-y^{\ast})(c+d)S
=-\frac{(aK_{1}+bK_{2})(aK_{2}+bK_{1})(c+d)}{(a+b)^{2}S}.
\end{align}
The eigenvalues are $\pm\lambda$ with
\begin{equation}
\lambda^{2}=\mathbf{X}_{y}\mathbf{Y}_{x}
=x^{\ast}(1-x^{\ast})\,y^{\ast}(1-y^{\ast})\,(a+b)(c+d)\,S^{2}.
\label{lambda2}
\end{equation}
Since the trace of $J$ is zero, this rest point can never be an attractor
or a repellor, in the classical or in the quantum game.

\begin{proposition}\label{prop:interior}
Suppose that the rest point (\ref{IntEq}) lies in the interior of the unit
square, $0<x^{\ast},y^{\ast}<1$. Then it is a saddle if $(a+b)(c+d)>0$ and a
centre if $(a+b)(c+d)<0$, for every initial state with $(a+b)(c+d)S\neq0$. In
particular, if an interior rest point exists in both the classical game and
a quantum form of it, its type is the same in both.
\end{proposition}

\begin{proof}
For an interior point $x^{\ast}(1-x^{\ast})>0$ and $y^{\ast}(1-y^{\ast})>0$,
and $S^{2}>0$, so by Eq.~(\ref{lambda2}) the sign of $\lambda^{2}$ is that of
$(a+b)(c+d)$, which does not depend on the $c_{ij}$. If $\lambda^{2}>0$ the
eigenvalues are real and of opposite sign (saddle); if $\lambda^{2}<0$ they
are purely imaginary, and the rest point is a centre by
Remark~\ref{rem:centre}.
\end{proof}

\begin{remark}\label{rem:centre}
Dividing the vector field (\ref{QRpEs}) by the positive function
$x(1-x)y(1-y)$ on the open square gives a system with the constant of
motion
$H=(cL_{2}+dL_{1})\ln x+(cL_{1}+dL_{2})\ln(1-x)
-(aK_{1}+bK_{2})\ln y-(aK_{2}+bK_{1})\ln(1-y)$,
as in the classical case \cite{schuster1,sigmund1}; since the rescaling
does not change orbits, $H$ is constant along the orbits of
Eqs.~(\ref{QRpEs}). Hence when $\lambda^{2}<0$ the interior rest
point is surrounded by closed orbits, so it is a true centre and not a
spiral.
\end{remark}

The classical value of Eq.~(\ref{lambda2}) is
$\lambda^{2}=abcd/[(a+b)(c+d)]$, and Proposition~\ref{prop:interior} shows
that quantization cannot turn a classical saddle into a centre, or a centre
into a saddle. What the MW transformation can do is move the rest point
(\ref{IntEq}) into or out of the unit square; it can also cause a
numerator or denominator in (\ref{IntEq}) to vanish, in which case
no isolated interior rest point exists (see Section~\ref{sec:degenerate}). For example, for $(a,b,c,d)=(1,2,-1,-3)$ the classical game has a
centre at $(x^{\ast},y^{\ast})=(3/4,1/3)$, surrounded by closed orbits.
For the initial state with
$(|c_{11}|^{2},|c_{12}|^{2},|c_{21}|^{2},|c_{22}|^{2})=(0.50,0.05,0.35,0.10)$
we have $S=0.2$ and $(x^{\ast},y^{\ast})=(11/8,\,5/12)$, which lies outside the
square. The quantum game therefore has no interior rest point, and hence no
periodic orbits, because in the plane a periodic orbit must enclose a rest point.
Quantization has destroyed the oscillatory behaviour of the classical game.

\subsection{Exact dynamics for parity-balanced initial distributions}
\label{sec:balanced}
An especially transparent parameter boundary is $S=0$, meaning that
$p_{11}+p_{22}=p_{12}+p_{21}=1/2$. It is a plane section of the
probability simplex, rather than a statement about entanglement. Here
$B=D=0$ and the dynamics decouples:
\begin{equation}
\dot x=A x(1-x),\qquad \dot y=C y(1-y).
\label{eq:decoupled}
\end{equation}
For initial frequencies $x_0,y_0\in(0,1)$, the exact solutions are
\begin{equation}
 x(t)=\frac{x_0e^{At}}{1-x_0+x_0e^{At}},\qquad
 y(t)=\frac{y_0e^{Ct}}{1-y_0+y_0e^{Ct}}.
\label{eq:logistic}
\end{equation}
If $A,C$ are nonzero, every interior trajectory converges to the
\emph{unique} attracting corner $(\mathbf 1_{A>0},\mathbf 1_{C>0})$,
where $\mathbf 1$ denotes an indicator function; the other three
corners are not attractors. If either coefficient vanishes, the
corresponding coordinate remains fixed, and the flow has continua of
rest points on suitable edges (or the entire square if both vanish).
In particular, $S=0$ does not imply a neutral system: it eliminates
cross-population frequency dependence while preserving directional
selection within each population whenever $A$ and $C$ are nonzero.
As with all results in this paper, the dynamics on this hyperplane
is determined entirely by the basis probabilities $p_{ij}$ and is
therefore equally reproduced by a separable classical mixture.

\subsection{Degenerate payoffs and boundary rest points}
\label{sec:degenerate}
The preceding interior classification assumes
$(a+b)(c+d)S\ne0$. To describe all other cases, write
\begin{align}
\dot x&=x(1-x)(A-By),\qquad
\dot y=y(1-y)(C-Dx),\notag\\
A&=aK_1+bK_2,\quad B=(a+b)S,\notag\\
C&=cL_2+dL_1,\quad D=(c+d)S.
\label{eq:ABCD}
\end{align}
An interior point is a rest point precisely if both equations
$A-By=0$ and $C-Dx=0$ hold. If $B=0$ and $A\ne0$, there is no
interior rest point; if $B=A=0$, the $x$ equation vanishes identically.
The analogous statements hold for $D$ and $C$. Consequently, when
$A=B=0$ and $D\ne0$, the interior equilibria (if any) comprise the
vertical line $x=C/D$; when $C=D=0$ and $B\ne0$, they comprise the
horizontal line $y=A/B$. If $A=B=C=D=0$, the entire square is fixed.
In all other degenerate cases there is no interior equilibrium.

In particular, $S=0$ forces $B=D=0$, so if $A,C$ are both nonzero
there are only the four vertices, whereas if either constant vanishes
one or both populations are stationary, with continua of fixed points
on the appropriate boundary or throughout the square. Nonvertex
boundary rest points also arise when $A=0$, $A=B$, $C=0$, or $C=D$,
as follows immediately by setting $x$ or $y$ equal to zero or one in
Eq.~(\ref{eq:ABCD}). The strict-NE criterion of
Proposition~\ref{prop:vertex} remains valid on all such parameter
boundaries: if a unilateral payoff difference vanishes at a vertex,
there is an adjacent edge of stationary points, so that vertex cannot
be asymptotically stable. Thus the statement ``strict NE if and only
if asymptotically stable'' holds also at nonhyperbolic vertices; the
linearization alone need not establish the converse in a generic
nonlinear system.

\section{Discussion}

In classical evolutionary game theory, attractors of a dynamics and ESSs
are usually studied in population models. Extending these ideas to quantum
settings requires assuming a population of individuals that have access
to quantum operations and to a shared, possibly entangled, initial state.
For two players the ESS concept retains a meaning if the term
``frequency'' is replaced by ``fraction of the total time''.
Two interacting quantum systems modelled as players in a game would then
raise questions of evolutionary stability and how it depends on the quantum
interaction. Proposition~\ref{prop:simulation} shows, however, that the
present MW protocol cannot serve as a model of such effects: any dynamics
it generates can equally be produced by classical shared randomness.
Addressing genuinely quantum evolutionary questions would require a
protocol with a richer operator set or non-diagonal measurements.

Proposition~\ref{prop:simulation} is a substantive limitation on
interpretation. All the phase portraits in Fig.~\ref{fig:phase} can be
reproduced by sampling the corresponding four-outcome classical
probability distributions before the players make their independent
identity/bit-flip choices. The observed changes are real properties of
the \emph{induced payoff matrices}, but none establishes that an
entangled state is necessary or that quantum coherence changes the
replicator dynamics. Distinguishing genuinely quantum effects would
require extending the experimental strategy/measurement protocol and
comparing it against an explicitly defined classical-correlation
benchmark. This also places the result in the context of earlier
analyses of classical correlation in quantum games
\cite{Brandenburger,Frackiewicz2020}.

The Marinatto--Weber construction \cite{marinatto} provides a
tractable parameterized family of evolutionary bimatrix games. As in classical ESS theory, pure strategies are combined with
classical mixing probabilities; the initial quantum state then supplies
additional parameters through which the stability of Nash equilibria and
attractors can be continuously varied. The ESS concept was earlier extended to
quantum games as a static concept \cite{iqbal}. Proposition~\ref{prop:vertex}
shows that in asymmetric games played in this scheme the static and the
dynamic characterizations coincide, exactly as in the classical theory.
This result provides a dynamical interpretation of earlier static
stability analyses where the same asymmetric $2\times2$ MW protocol is
used; it does not extend the claim to the distinct symmetric-game or
higher-dimensional models considered in parts of
Refs.~\cite{iqbal,iqbal1,iqbal2,iqbal3,iqbal4}.
It also limits what quantization can achieve. The quantum game is again a
bi-matrix game, and its payoff matrices are linear in the classical ones
with coefficients given by the probabilities $|c_{ij}|^{2}$.
Proposition~\ref{prop:regions} makes the resulting stability changes
explicit as polyhedral regions of the probability simplex. For this
reason, any feature of the replicator dynamics that is shared by all
bi-matrix games is inherited unchanged. Changes arise only in features
that depend on the particular payoffs: which vertices are strict NE, and
whether an interior rest point exists.

An important way in which evolutionary game theory differs from classical
game theory is that it does not need rational decision makers \cite{Bomze}.
In evolutionary game theory an individual's strategy is an inherited trait,
and a population is an abstract collection of interacting individuals with
genetically determined strategies. Likewise, in our extension of these ideas
to quantum games, no rationality is attributed to the agents whose actions
are quantum mechanical. Such physical interpretations would additionally need a microscopic
mechanism for payoff-dependent reproduction or updating and for repeated
preparation of the shared state; these mechanisms are not derived here.

Matrix games provide one of two perspectives on the outcome of evolution;
the other is provided by optimization models \cite{meszena}. In
optimization models selection is frequency independent, evolution is
pictured as hill climbing, and the outcome maximizes fitness. Evolutionary
optimization underlies evolutionary and genetic algorithms. The two
perspectives are not in direct contradiction but offer different views of
the evolutionary process \cite{meszena}, and the role of quantum effects in
the optimization perspective remains to be explored.

\section{Conclusion}
For the asymmetric $2\times2$ Marinatto--Weber protocol, the induced
payoffs form a classically simulable family of bi-matrix games. We have
shown directly that a pure-strategy vertex of the two-population
replicator dynamics is an asymptotically stable equilibrium precisely
when it is a strict Nash equilibrium, and hence an ESS pair in the
asymmetric-game sense. Varying the shared initial-state probabilities
can exchange stable and unstable vertices, as illustrated by two
explicit examples with calculated transition thresholds. More generally,
the stable-vertex regions are polyhedral subsets of the probability
simplex (Proposition~\ref{prop:regions}), with a quantitative
robustness bound. A nondegenerate interior equilibrium, when it
exists, is a saddle or a true centre according to the sign of the
original payoff cross-differences; changing the initial distribution
can move it across the boundary but not change its type. The parity-balanced family has an explicit logistic solution
(Section~\ref{sec:balanced}); the additional
degenerate cases are characterized in Section~\ref{sec:degenerate}:
they include lines and continua of rest points on the boundary or in the interior.
Crucially, all these results are unchanged if the pure quantum initial
state is replaced with a separable diagonal state with the same basis
probabilities. Claims of quantum-specific evolutionary effects therefore
require protocols or observables that go beyond those analysed here.


\begin{thebibliography}{99}

\bibitem{Piotrowski}
E.~W.~Piotrowski and J.~S{\l}adkowski, An invitation to quantum game
theory, Int.\ J.\ Theor.\ Phys.\ \textbf{42}, 1089 (2003),
arXiv:quant-ph/0211191.

\bibitem{Piotrowski2}
E.~W.~Piotrowski and J.~S{\l}adkowski, The next stage: quantum game
theory, arXiv:quant-ph/0308027 (2003).

\bibitem{Flitney}
A.~P.~Flitney and D.~Abbott, An introduction to quantum game theory,
Fluct.\ Noise Lett.\ \textbf{2}, R175 (2002), arXiv:quant-ph/0208069.

\bibitem{Neumann}
J.~von Neumann and O.~Morgenstern, \emph{Theory of Games and Economic
Behavior} (Princeton University Press, Princeton, 1944; 3rd ed., 1953).

\bibitem{Weibull}
J.~W.~Weibull, \emph{Evolutionary Game Theory} (MIT Press, Cambridge, MA,
1995).

\bibitem{Smith}
J.~Maynard Smith and G.~R.~Price, The logic of animal conflict,
Nature \textbf{246}, 15 (1973).

\bibitem{MaynardSmith}
J.~Maynard Smith, \emph{Evolution and the Theory of Games}
(Cambridge University Press, Cambridge, 1982).

\bibitem{iqbal}
A.~Iqbal and A.~H.~Toor, Evolutionarily stable strategies in quantum
games, Phys.\ Lett.\ A \textbf{280}, 249 (2001).

\bibitem{iqbal1}
A.~Iqbal and A.~H.~Toor, Entanglement and dynamic stability of Nash
equilibria in a symmetric quantum game, Phys.\ Lett.\ A \textbf{286}, 245
(2001).

\bibitem{iqbal2}
A.~Iqbal and A.~H.~Toor, Darwinism in quantum systems?, Phys.\ Lett.\ A
\textbf{294}, 261 (2002).

\bibitem{iqbal3}
A.~Iqbal and A.~H.~Toor, Quantum mechanics gives stability to a Nash
equilibrium, Phys.\ Rev.\ A \textbf{65}, 022306 (2002).

\bibitem{iqbal4}
A.~Iqbal and A.~H.~Toor, Stability of mixed Nash equilibria in symmetric
quantum games, Commun.\ Theor.\ Phys.\ \textbf{42}, 335 (2004),
arXiv:quant-ph/0106056.

\bibitem{marinatto}
L.~Marinatto and T.~Weber, A quantum approach to static games of complete
information, Phys.\ Lett.\ A \textbf{272}, 291 (2000),
arXiv:quant-ph/0004081.

\bibitem{cressman}
R.~Cressman, \emph{The Stability Concept of Evolutionary Game Theory},
Lecture Notes in Biomathematics Vol.~94 (Springer, Berlin, 1992).

\bibitem{TaylorJonker}
P.~D.~Taylor and L.~B.~Jonker, Evolutionarily stable strategies and game
dynamics, Math.\ Biosci.\ \textbf{40}, 145 (1978).

\bibitem{HSS}
J.~Hofbauer, P.~Schuster and K.~Sigmund, A note on evolutionary stable
strategies and game dynamics, J.\ Theor.\ Biol.\ \textbf{81}, 609 (1979).

\bibitem{Zeeman}
E.~C.~Zeeman, Population dynamics from game theory, in \emph{Global Theory
of Dynamical Systems}, edited by Z.~Nitecki and C.~Robinson, Lecture Notes
in Mathematics Vol.~819 (Springer, Berlin, 1980), p.~471.

\bibitem{Selten}
R.~Selten, A note on evolutionarily stable strategies in asymmetric animal
conflicts, J.\ Theor.\ Biol.\ \textbf{84}, 93 (1980).

\bibitem{RitzbergerWeibull}
K.~Ritzberger and J.~W.~Weibull, Evolutionary selection in normal-form
games, Econometrica \textbf{63}, 1371 (1995).

\bibitem{HS2003}
J.~Hofbauer and K.~Sigmund, Evolutionary game dynamics, Bull.\ Am.\ Math.\
Soc.\ \textbf{40}, 479 (2003).

\bibitem{Bomze}
I.~M.~Bomze and B.~M.~P\"{o}tscher, \emph{Game Theoretical Foundations of
Evolutionary Stability}, Lecture Notes in Economics and Mathematical
Systems Vol.~324 (Springer, Berlin, 1989).

\bibitem{sigmund1}
J.~Hofbauer and K.~Sigmund, \emph{Evolutionary Games and Population
Dynamics} (Cambridge University Press, Cambridge, 1998).

\bibitem{schuster1}
P.~Schuster and K.~Sigmund, Coyness, philandering and stable strategies,
Anim.\ Behav.\ \textbf{29}, 186 (1981).

\bibitem{Hirsch}
M.~W.~Hirsch and S.~Smale, \emph{Differential Equations, Dynamical
Systems, and Linear Algebra} (Academic Press, New York, 1974).

\bibitem{Broom}
M.~Broom, Patterns of evolutionarily stable strategies: the maximal
pattern conjecture revisited, J.\ Math.\ Biol.\ \textbf{40}, 406 (2000).

\bibitem{meszena}
G.~Mesz\'{e}na, \'{E}.~Kisdi, U.~Dieckmann, S.~A.~H.~Geritz and
J.~A.~J.~Metz, Evolutionary optimisation models and matrix games in the
unified perspective of adaptive dynamics, Selection \textbf{2}, 193 (2001).

\bibitem{CressmanTao}
R.~Cressman and Y.~Tao, The replicator equation and other game dynamics,
Proc.\ Natl.\ Acad.\ Sci.\ USA \textbf{111}, 10810--10817 (2014),
doi:10.1073/pnas.1400823111.

\bibitem{Frackiewicz}
P.~Fr\c{a}ckiewicz, A comment on the generalization of the Marinatto--Weber
quantum game scheme, Acta Phys. Pol. B \textbf{44}, 29--33 (2013),
doi:10.5506/APhysPolB.44.29; arXiv:1303.4019.

\bibitem{Brandenburger}
A.~Brandenburger, The relationship between quantum and classical
correlation in games, Games Econ.\ Behav.\ \textbf{69}, 175--183 (2010),
doi:10.1016/j.geb.2009.10.009.

\bibitem{Frackiewicz2020}
P.~Fr\c{a}ckiewicz, On correlated equilibria in Marinatto--Weber type
quantum games, Appl.\ Sci.\ \textbf{10}, 9003 (2020),
doi:10.3390/app10249003.

\end{thebibliography}
\end{document}